\documentclass[letterpaper,12pt]{article}
\usepackage{graphicx}
\usepackage{fancyvrb}
\usepackage{tikz}
\usepackage{verbatim}

\usepackage[bottom=1in,top=1in,left=1.25in,right=1.25in]{geometry}                		
\usepackage{graphicx}				
\usepackage{amssymb}

\usepackage[ruled,vlined]{algorithm2e}
\usetikzlibrary{arrows}
\usepackage{alltt}
\usepackage[T1]{fontenc}
\usepackage[utf8]{inputenc}
\usepackage{indentfirst}
\usepackage[longnamesfirst]{natbib} 
 \bibpunct{(}{)}{;}{a}{}{,} 
\usepackage{changepage}
\usepackage{setspace}
\usepackage{booktabs} 
\usepackage{rotating} 
\usepackage{amsmath}
\usepackage{multirow}
\usepackage{color}
\usepackage{dcolumn}
\usepackage{comment}
\newcolumntype{d}[1]{D{.}{\cdot}{#1}}
\newcolumntype{.}{D{.}{.}{-1}}
\newcolumntype{3}{D{.}{.}{3}}
\newcolumntype{4}{D{.}{.}{4}}
\newcolumntype{5}{D{.}{.}{5}}
\usepackage{float}
\usepackage[small]{caption}
\usepackage{subfig}
\usepackage{fancyhdr}
\usepackage{times}
\usepackage{mathptmx}
\def\citeapos#1{\citeauthor{#1}'s (\citeyear{#1})}

\newcommand{\R}{\textsf{R}\space} 

\newcommand{\bZ}{\mathbf{Z}}

\newcommand{\bY}{\mathbf{Y}}
\newcommand{\bx}{\mathbf{x}}

\graphicspath{{.}{../figures/}}
\title{\vspace{-2cm} Models, Methods and Network Topology: \\ Experimental Design for the Study of Interference\footnote{The authors would like to thank Carlisle Rainey for providing a detailed review of an early draft of this paper. The authors gratefully acknowledge the role of the Statistical and Applied Mathematical Sciences Institute in creating the environment that allowed this work to be produced. This work was supported in part by NSF grants SES-1558661, SES-1637089, SES-1619644, SES-0752986, and CISE-1320219. Any opinions, findings, and conclusions or rec- ommendations are those of the authors and do not necessarily reflect those of the sponsors.}}

\date{\today}

\author{ Jake Bowers \thanks{\footnotesize{Associate Professor, Departments of
Political Science \& Statistics, University of Illinois at Urbana--Champaign,
jwbowers@illinois.edu.}} \and Bruce A. Desmarais
\thanks{\footnotesize{Associate Professor, Department of Political Science,
Pennsylvania State University, bdesmarais@psu.edu.}}\and Mark Frederickson
\thanks{\footnotesize{PhD Student, Departments of Political Science and
Statistics, University of Illinois at Urbana-Champaign, }} \and  Nahomi Ichino
\thanks{\footnotesize{Assistant Professor, Department of Political Science,
University of Michigan}} \and  Hsuan-Wei Lee
\thanks{\footnotesize{Post-doctoral Student, Department of Sociology,
University of Nebraska-Lincoln}} \and Simi Wang \thanks{\footnotesize{PhD Student, Department of Applied Mathematics, University of North Carolina at Chapel Hill}} \footnote{Authors listed alphabetically.}} \date{\today}

\begin{document}
\begin{titlepage}
\maketitle


\begin{abstract}

 \noindent How should a network experiment be designed to achieve high
 statistical power? Experimental treatments on networks may spread.
 Randomizing assignment of treatment to nodes enhances learning about the
 counterfactual causal effects of a social network experiment and also
 requires new methodology
 \citep[ex.][]{aronow2017interference,Bowers:2013,Toulis:2013}. In this paper
 we show that the way in which a treatment propagates across a social network
 affects the statistical power of an experimental design. As such, prior
 information regarding treatment propagation should be incorporated into the
 experimental design.  Our findings justify reconsideration of standard
 practice in circumstances where units are presumed to be independent even in
 simple experiments: information about treatment effects is \emph{not}
 maximized when we assign half the units to treatment and half to control. We
 also present an example in which statistical power depends on the extent to
 which the network degree of nodes is correlated with treatment assignment
 probability.   We recommend that researchers think carefully about the
 underlying treatment propagation model motivating their study in designing an
 experiment on a network. \\~\\

\end{abstract}
\thispagestyle{empty}
\end{titlepage}

\doublespacing
\section{Introduction}

We consider the problem of designing experiments to causally identify propagation on networks.
In a simple experiment on \emph{independent} units with complete randomization to two
treatment arms, it is often assumed that one should assign half of the experimental pool to treatment and half
to control \citep{gerber2012field}.\footnote{Technically speaking, the 50/50 treatment allocation
is optimal for precision when randomization is complete at the unit-level and
outcomes have equal variance in both treated and control groups.} When treatment given to one unit
may affect another unit, however, we show (in a simulation study using a
realistic network and realistic model of network treatment propagation) that
it may be better to assign \emph{less} than half of the pool to treatment from
the perspective of statistical efficiency. The intuition is simple: if treatment spreads rapidly across a network, then comparisons of outcomes between treated and control units will become very small or even vanish as the control units to which the treatment spread will act just like treated units. Thus, one might field a very effective experiment, perhaps an experiment in which controls race to get access to the treatment or treated units spread the information or other active ingredient far and wide, but be unable to detect effects if everyone in the whole network reveals the same outcome whether or not they were assigned to treatment. The simulations that we show here confirm this intuition, but also reveal a trade-off between ability to detect the direct effects of treatment assignment on the units initially assigned to treatment and the ability to detect the indirect or network mediated effects of the treatment as it propagates to control units. One point that we emphasize in this paper is that the way in which a treatment propagates matters a great deal as we think about how to design experiments on networks.

In fields across the social and physical sciences, there is considerable and growing interest in understanding how features propagate over the vertices (i.e., nodes) in a graph (i.e., network) via the graph topology. Furthermore, precise questions about causal peer, spillover and propagation effects are becoming more common. Recent theoretical developments highlight the barriers to the identification of causal peer/contagion effects in networks with non-randomized, or observational, data \citep{Lyons2011,Shalizi:2011}. Several recent papers have employed randomized experimental designs to facilitate the identification of causal peer effects \citep{Aral:2011,Ostrovsky:2011,Bapna:2015,Bond:2012, Ichino:2012,nickerson2008voting}. For example, \citet{Ichino:2012} conduct a field experiment during a national election in Ghana to gauge how voter registration responds to the placement of election monitors at registration workstations---an effect that is hypothesized to spread geographically through the road network.

Recent methodological work enables scholars to make statistical inferences about
peer effects or global average effects when the topology of a network is known
\citep{Bowers:2013,aronow2017interference, eckles2017design,Toulis:2013}.\footnote{For now, we set to
  the side the work on identifying how much of a total average effect can be
  attributed to mechanisms other than direct treatment assignment --- for
  example, the work on spillovers and indirect effects
  \citep{sinclair2012detecting,sinclair33, nickerson2008voting,
    nickerson2011social, hudgens2008toward, sobel2006randomized,
    tchetgen2010causal, vanderweele2008ignorability, vanderweele2010direct,
    vanderweele2011components, vanderweele2012mapping,
    vanderweele2011bounding, vanderweele2011effect, miguel2004worms,
    chen2010technology, Ichino:2012}.}
As the ability to pose questions of spillover has increased, researchers have begun to address how well these methods work, particularly with respect to statistical efficiency.
\citet{eckles2017design} show that a graph
  cluster randomization design --- where groups of nodes are randomized to
  treatment together --- reduces bias in estimates of global average
  treatment effects with relatively little cost in terms of statistical power.
  \citet{baird2017optimal} derive the efficiency calculations for
estimates of average spillover effects for randomization designs in which isolated
groups of nodes are randomized first to a saturation proportion --- the
proportion of units within the group to be randomized to treatment --- and
then within group randomization proceeds according to the first level
randomization. \citet{hirano2010design} derive efficiency calculations regarding cluster-wise and within-cluster treatment proportions for estimates of direct and indirect effects in two-level cluster randomization designs. These approaches answer important questions about particular designs; however, there is still a need to address  how to
\emph{design randomization schemes} to increase the statistical power to
detect specific forms of network mediated peer
effects.

In this project we consider the performance of different randomization designs using the methods of \citet{Bowers:2013} and \citet{aronow2017interference} under different models of propagation. Each of the methods we consider depends upon a typology of exposure conditions based on the treatment status of each node and the topology of the graph. For example, a node could be treated directly by an experimenter, isolated from treatment (i.e., several hops away from any treated nodes) or exposed to the treatment at one degree of separation by virtue of the network relationship --- without control by the experimenter. The performance of randomized experimental designs on networks depends on (1) the exposure conditions of theoretical interest (say, direct treatment versus indirect treatment; or more generally some propagation flow parameter), (2) the topology of the network, (3) the ways in which the propagation model affects nodes in each exposure condition, and (4) the exposure condition distribution as determined by the randomization design.\footnote{We direct readers to \citet{Basse2015} for a methodological investigation similar to ours. They consider the problem of designing a randomized experiment to minimize estimation error when outcomes are correlated on a network. Their focus is, however, on estimating the direct effects of treatment, not on identifying indirect or propagation effects.}

To anchor our interest in interference, consider \citeapos{coppock2014information} recent replication of \citet{butler2011can}.  \citet{butler2011can} run a field experiment that is focused on a special session of the New Mexico legislature that was called to consider a specific budgetary question. The field experiment was designed to test the influence of providing information about constituents' preferences on legislators' votes.  Constituents across the state were first surveyed on the budget question on which their legislators would be voting. Butler and Nickerson sent district-specific results to randomly selected members of the legislature.  They found that providing information about constituents' preferences  shifted legislators' votes in the direction of those preferences. \citet[pp.  159--160]{coppock2014information} notes that, \begin{quotation} \noindent ``The estimates of responsiveness recovered by Butler and Nickerson (2011) rely on an assumption of non-interference (Cox 1958; Rubin 1980): Legislators respond only to their own treatment status and not to the treatment status of others. This assumption requires that legislators not share treatment information with one another, which is at odds with the observation by Kingdon (1973, p. 6) that legislatures are information-sharing networks.'' \end{quotation} In replicating \citet{butler2011can}, \citet{coppock2014information} specifies a model for the propagation of effects that spread through a network between legislators defined by ideological similarity.  Accounting for the fact that the treatment assigned to one legislator had effects on other legislators, \citet{coppock2014information} estimates that the experiment shifted nearly twice as many votes in the legislature as was originally estimated by \citet{butler2011can}.\footnote{\citet{coppock2016information} later shows that the test statistic and research design was underpowered to detect this effect.}

In what follows, we study the problem of causal inference given treatment propagation in the context of a fixed graph topology and a single round of randomized treatment and by a single round of response measurement. We review methods that have been proposed in the literature for analyzing single-round (pre versus post), fixed graph experimental data; and also review the substantive experimental applications that have used such designs. We then conduct a simulation study motivated by the registration monitor randomization in \citet{Ichino:2012}, using the Ghanaian network of roads between voter registration stations as  a realistic moderate sized graph.\footnote{ \citet{Ichino:2012} did not use the road network in their paper, but instead focused on estimating average spillover effects within radii of 5km and 10km following the multi-level experimental design of \citet{sinclair2012detecting}. We use the road network to provide us with a realistic network for use in our simulations studying the power of different randomization allocation plans.} In the simulation study, we consider the performance of alternative experimental designs that vary the treatment probability: the number of nodes assigned to initial treatment, who is treated: the association between treatment probability and node degree (i.e., a node's number of ties), and how they are treated: different parameterizations of the propagation model.

\subsection{Statistical Inference for Propagated Causal Effects}

We consider two general approaches to statistical inference about causal effects when those effects may propagate through a network. The flexible approach developed by \citet{Bowers:2013} is a hypothesis testing framework designed to evaluate whether differences between the treatment and control groups are more effectively characterized by one model of treatment effects, which can include propagation effects, than another model. \citet{Bowers:2013} focus on a natural sharp null model of no treatment effects (i.e., stochastic equivalence across all experimental conditions). The null distribution is derived exactly or generated approximately through repeated computations of the test statistic using permutations in which the treatment vector is re-randomized according to the experimental design, and the hypothesized effects of the propagation model are removed. There are two highly appealing properties of this approach. First, any test statistic, including general distributional comparisons such as the Kolmogorov-Smirnov (KS) test statistic, can be used to evaluate the differences between treatment and control. Second, the approach can accommodate any model of treatment effects on a network, as the methodology does not require any form of independence assumption or the derivation of an estimator for the model parameters.

The methods developed by \citet{aronow2017interference} and \citet{Toulis:2013} compliment those proposed by \citet{Bowers:2013} in that they propose methods for {\em estimating} average causal peer effects. \citet{aronow2017interference} develops randomization-based methods and  \citet{Toulis:2013} develops both randomization and model-based approaches to estimating causal peer effects.  In both \citet{aronow2017interference} and \citet{Toulis:2013}, the target estimate is the average difference between nodes in different network/treatment exposure conditions. \citet{aronow2017interference} do not stipulate a constrained set of conditions, but present methods that can be applied to any partition of nodes into network/treatment exposure conditions. They present an example in which nodes in a graph are directly treated and assume that treatment can only propagate one degree, which results in four conditions: {\em control}, which are nodes that are not directly assigned treatment and are not tied to any treated nodes; {\em direct}, which are nodes that are treated and not tied to any treated nodes, {\em direct $+$ indirect}, which are nodes that are directly treated and are tied to treated nodes; and {\em isolated direct}, which are nodes that are untreated and are tied to treated nodes.  \citet{Toulis:2013}  define $k$-level treatment of a unit as (1) a unit not receiving direct treatment, and (2) having exactly $k$ directly treated neighbors.  A $k$-level control is any vertex with at least $k$ neighbors who did not (1) receive direct treatment and (2) is not connected to any vertices who were directly treated. These approaches assume that the researcher is interested in specific comparisons of averages and require that the researcher articulate mechanisms by which the probability of exposure to treatment may differ across units. When networks are fixed prior to treatment and the randomization mechanism known, the probability of exposure can be directly computed.

Both \citet{aronow2017interference} and \citet{Toulis:2013} recognize the unique challenges that arise in the context of inference regarding response to network/treatment exposure conditions. The limitations are based in the topology of the graph. Since most exposure conditions of interest in the context of interference involve the position of a node in a network, under most randomization designs (e.g., uniform random assignment to treatment), each node is not equally likely to be assigned to each exposure condition. Take the example of $2$-level exposure in the framework of \citet{Toulis:2013}. A node with only one partner would have zero probability of being assigned to the $2$-level treatment group.   \citet{aronow2017interference} do not discuss this issue at length, but imply a limitation in the derivation of their Horvitz-Thompson type estimators \citep{Horvitz1952}. The estimators they define require that the analyst be able to calculate the probability $\pi_i(d_k)$, the probability that node $i$ is in exposure condition $d_k$ and that $0 < \pi_i(d_k) < 1$ for each node $i$. This means that the framework proposed by \citet{aronow2017interference} cannot be applied to the comparison of exposure conditions to which all nodes cannot be assigned.   \citet{Toulis:2013}  are more explicit in their discussion of this limitation. They define a causally valid randomization design to be one in which at least one node is $k$-level treated and one is $k$-level controlled.

In the analysis that follows, we consider the \citet{aronow2017interference} and \citet{Bowers:2013} approaches to inference with experiments on networks. The methods proposed by \citet{Toulis:2013} are very similar to those of \citet{aronow2017interference}, and their concepts of $k$-level treated and $k$-level controlled can be seen as special cases of the exposure conditions defined in \citet{aronow2017interference}. Furthermore, the methods of \citet{aronow2017interference} have the added advantage of adjusting the treatment effect estimates (and variance estimates) for the unequal exposure condition probabilities. These Horvitz-Thompson type adjustments will correct for any associations between exposure condition probabilities and potential outcomes (e.g., higher degree nodes may exhibit higher baseline response values {\em and} be more likely to be indirectly exposed to treatment through propagation).

\section{Design as a Function of Graph Topology}

\citet{Walker:2014} reviews several applications of experiments in networked environments --- including studies that are not focused on propagation --- and outline several of the fundamental challenges to designing experiments on networks. They summarize the problem of design in experiments on networks succinctly (p. 1949):

\begin{quotation}\noindent``The natural connectivity of our world does not only
present a challenge to the conventional paradigm of
experimental design, but also reveals opportunities to
leverage connectivity through the creation of novel
treatment mechanisms that incorporate both experimental
subjects and the connections between them.''
\end{quotation}

The practical implication of the dependence between subjects via the network is that efficient experimental designs will account for graph topology. The treatment assignment algorithms presented in \citet{Toulis:2013} render a clear picture of how the importance of network structure can complicate design. Considering the problem of assuring that sufficient numbers of vertices end up in the $k$-treated and $k$-controlled designs, \citet{Toulis:2013} present sequential randomization designs that assure that fixed numbers of vertices are assigned to the groups under comparison: for example, if a researcher desires to know the effect of the treatment on nodes having 2 directly connected neighbors in the network, a design should ensure enough nodes with degree 2 assigned treatment versus not are assigned treatment. Though powerful in their ability to control the distribution of vertices across exposure conditions, the complex sequential randomization algorithms proposed by \citet{Toulis:2013} make closed-form calculation of the probability of exposure condition assignment intractable in most cases, which may be why they do not derive their estimators using Horvitz-Thompson adjustments such as those in \citet{aronow2017interference}. An example of a non-sequential randomization design for which it is straightforward to derive the Horvitz-Thompson weights is one in which the probability of treatment is biased with respect to vertex degree (e.g., disproportionately treating higher degree vertices). To provide an intuitive example regarding why it might be advantageous to treat high degree vertices at a greater rate than low degree vertices; suppose the researcher is interested in comparing nodes isolated from treatment (e.g., more than two degrees from any directly treated node) to nodes that are adjacent to a treated node, but are not directly treated. Each node that is directly treated is removed from both exposure conditions of interest, so there is an incentive to treat a small proportion of nodes. However, if too few nodes are treated, there will be too few nodes in the adjacent-to-treated condition. By focusing the treatment on higher degree nodes, it takes fewer directly treated nodes to accomplish a sizable sample of adjacent-to-treated nodes. Depending upon the structure of the network and the mechanism by which treatment can propagate, there may be a considerable gain in statistical power from biasing treatment towards high degree nodes, as compared to uniform assignment to treatment.

\begin{figure}[!ht]
\begin{center}
\begin{tabular}{cc} \vspace{-.3cm}
(a) Example $3 \times 4$ Grid Network & (b) Exposure condition distributions \\ \vspace{-.3cm}
\includegraphics[width=.4\textwidth, trim = 1cm .25cm 1cm 1.5cm, clip=true]{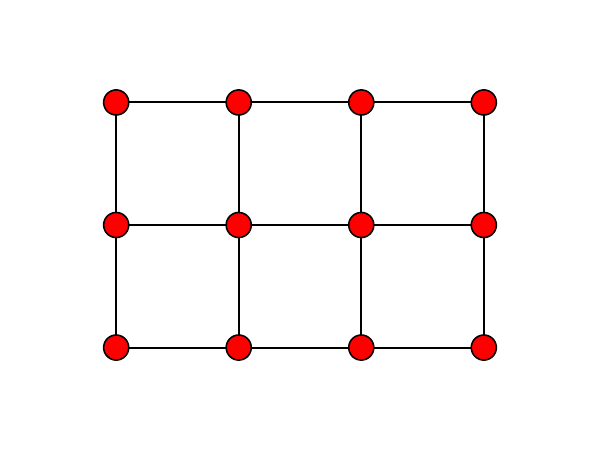} &\includegraphics[width=.5\textwidth, trim = 0cm 1cm 1cm 1cm, clip=true]{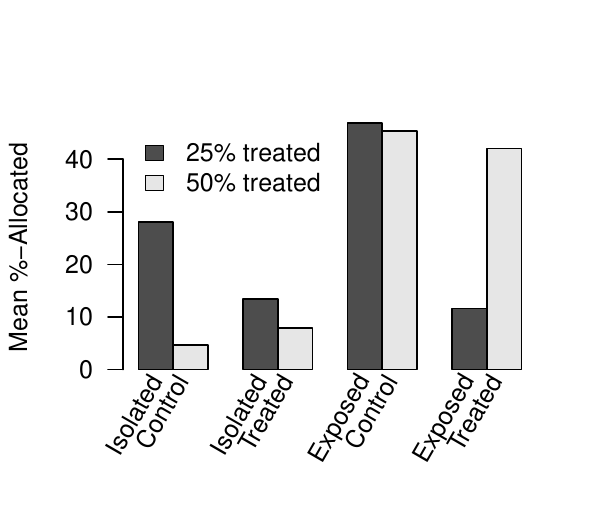}  \\
\end{tabular}
\end{center}
\caption{Example illustrating how simple random treatment assignment on a network can lead to very different allocations of nodes to network exposure conditions. Panel (a) gives an example grid network of twelve nodes. Panel (b) gives the average allocation across exposure conditions given simple random assignment to treatment of both three (i.e., 25\%) and six (i.e., 25\%) nodes. }
\label{fig:gridexample}
\end{figure}

Consider a simple example that illuminates how the ability to draw comparisons in experiments on networks can depend significantly on design decisions. In this example, an experiment is conducted by allocating a binary treatment in a network of twelve nodes connected on a $3 \times 4$ grid (illustrated in Figure \ref{fig:gridexample} (a)). Suppose the experimenter is planning to draw some comparisons across four exposure conditions of nodes:

\begin{itemize}
\item {\em Isolated Control}: Control nodes that are not adjacent to any treated nodes.
\item {\em Isolated Treated}: Treatment nodes that are not connected to any treated nodes.
\item {\em Exposed Control}: Control nodes that are adjacent to at least one treated node.
\item {\em Exposed Treated}: Treated nodes that are adjacent to at least one treated node.
\end{itemize}

Now consider two designs for assigning nodes to treatment. In each design, the set of treated nodes is selected uniformly at random from the set of twelve nodes. In the first design, treatment is assigned to 25\% of the nodes (i.e., three). In the second design, treatment is assigned to 50\% of the nodes (i.e., six). In Figure \ref{fig:gridexample} (b) we present the average percentage of nodes allocated to each of the exposure conditions. Generally speaking, the allocation to the four conditions differs dramatically between the two designs. More specifically, under the design in which 50\% of the nodes are allocated to treatment, fewer than 15\% of the nodes are isolated from treatment, on average.  If the experimenter is planning to conduct analysis that depends on the number of nodes that are isolated from treatment, the design in which treatment is assigned to 50\% of the nodes will likely result in very imprecise estimates and/or low power statistical tests. 

This simple example illustrates two points upon which we build throughout the paper. First, the power of a network experiment depends, in complex ways, on the structure of the network, the treatment assignment distribution, and the effects of the treatment---both direct and indirect. Second, if the precision in the estimates/analysis depends upon the number of subjects that are isolated from treatment, following the common practice of allocating half the experimental subjects to treatment may result in very low power/precision. 

\subsection{Design is Dependent on a Model of Propagation}

Let $Z_i = 1$ if subject $i$ is assigned to the treatment condition and $Z_i = 0$ if $i$ is assigned to control. In the classical experimental framework, a counterfactual causal effect for a person $i$ is defined through the comparison of that person's outcome, $Y_i(Z_i=1)$ under one treatment to that same person's outcome under a different treatment $Y_i(Z_i=0)$, given at the same moment in time \citep{neyman:1923,Rubin:1980qy,Holland:1986ly}. If potential outcomes under different treatments differ, $Y_i(Z_i) \ne Y_i(Z'_i)$ for $Z_i \ne Z'_i$, we say the treatment had an effect. Implicit in this comparison for a single person is the presumption that subject $i$'s outcome only depends on $i$'s treatment status $Z_i$ and not on any other $Z_j$ --- that there is no interference between the treatment or outcomes of person $i$ and those of any other person. Such implicit assumptions are incompatible with the goals of researchers seeking to understand the important causal role played by networks in real world processes. Analysis of experiments on social networks cannot assume away interference, especially if a scholar desires to learn about propagation processes. However, once the assumption of no interference is relaxed, a theoretical model of interference represents a valuable tool to guide the search for propagation's footprint and to calibrate the experimental design. One method to deal with interference would be to specify the potential outcomes for each unit as a function of the entire treatment assignment to all nodes \citep{Bowers:2013}. Such a method is powerful, but it can be theoretically demanding to specify a model for each unit's outcome. As \citet{aronow2017estimating} show, it is not strictly necessary to assume a theoretical model of interference in order to identify interference effects by instead looking at broad classes of interference. Nevertheless, in both scenarios, a theoretical model of interference is needed to identify the interference effects for which the researcher intends to test, which, in turn, can inform the design of the experiment.

While having differing statistical motivations, the methods of \citet{aronow2017interference} and \citet{Bowers:2013} both require that the researcher draw upon a theoretical model of interference to properly analyze the outcome $Y$. In \citet{aronow2017interference}, a model of interference must be used to identify the exposure conditions (e.g., adjacent to a treated unit, 2 degrees of separation from a treated unit, etc.) to be compared in the study. The model used with the methods of \citet{aronow2017interference} need not provide precise predictions about to which nodes the treatment will propagate and how, but the model must be complete enough to identify the groups of nodes for which different potential outcomes will be observed under a given direct treatment assignment regime. For the methods proposed by \citet{Bowers:2013}, a precise analytic model of interference must be specified. The approach adjusts the observed data by removing effect of the treatment as specified by a parametric model of effects, including both direct and indirect effects. If the model captures the true treatment effects, outcome and treatment assignment should be statistically independent for any test statistic. This statistical independence can be tested by permuting the treatment assignment labels, computing a test statistic for each assignment, and comparing the test statistic of the adjusted data to the distribution of test statistics. As the adjusted data must be constructed from a model, it is not possible to use the methods of \citet{Bowers:2013} without specifying the precise role of interference in the study.

The need to specify a model of propagation in order to identify an efficient experimental design leads us to question where researchers might start in developing such models. There is a vibrant literature, primarily in the fields of physics and applied mathematics, on graph dynamics, that provides several excellent starting points for analytical models of propagation. These models include the susceptible-infected-recovered disease epidemic models \citep{Kermack_Mathematical_1927, Anderson_Population_1982, Hethcote_The_2000, Daley_Epidemic_2005},  the Bass Diffusion Model \citep{Bass_Why_1994, Lenk_New_1990} and the Hopfield network \citep{Hopfield_Neural_1982} and the voter model \citep{Clifford_A_1973, Liggett_Stochastic_1997, Durrett_Random_1991}. In the simulation study that follows, we present and use a variant of the Ising model, a model that contains several of those mentioned above as special cases \citep{Gallavotti:1999}.  The Ising model is a general formulation for stochastic binary-state dynamics. The classic Ising model has been used quite widely to characterize opinion dynamics \citep[e.g., ][]{vazquez2003constrained,fortunato2005sznajd,sousa2008effects,biswas2009model}.

\section{Simulation Study}

In what follows we conduct a simulation study in which we evaluate the statistical power of the methods proposed by \citet{aronow2017interference} and \citet{Bowers:2013}. The objective of this simulation study is to demonstrate that the statistical power of these procedures depends upon design parameters that are intuitively meaningful in the context of propagation, and that power is maximized at design parameters that differ considerably from those commonly used in experiments without interference --- random uniform division of the sample into half control and half treatment.\footnote{See \citet[Chap 3]{gerber2012field} for one example of a discussion about why, in general, a half/half treatment/control split maximizes power.} It is outside the scope of the current study to compare the merits of the methods proposed by \citet{aronow2017interference} with those proposed by \citet{Bowers:2013}: further, we see the two approaches as complementary in the same way that hypothesis testing and estimation complement one another.  We hope that our simulation will illustrate the importance and feasibility of simulation analysis for parameterizing designs for studies of propagation in networks. Throughout this simulation we use the Ghana voter registration station network from \citet{Ichino:2012} as our example network. We create a network from the map of roads connecting the registration stations.  Two registration stations are considered to be tied if they are within 20km of each other on the road network. This results in a network with 868 vertices, and a density of 2.2\%. The network is depicted in Figure \ref{fig:networks}.\footnote{We depart from the details and substantive aims of \citet{Ichino:2012} in that we replace a road network with a graph that creates direct connections between nodes. This enables us focus on network propagation rather than, say, the movements of actual human agents on road, or on the specifics of the study of the propagation of voter registration fraud in Ghana.}

\subsection{Simulation Parameters and Definitions}

We denote the treatment assigned to vertex $i$ as $Z_i \in \{0,1\}$ and the vector of treatment assignments to all vertices as $\bZ= (Z_1,\ldots,Z_n )^T$.  The fixed potential outcome of vertex $i$ that would be shown under treatment vector $\bZ$ is denoted $Y_i(\bZ)$. Under the sharp null of no treatment effect, all potential outcomes are equal; $Y_i(\bZ)=Y_i(\bZ')=Y_i(\{Z_i=1,\bZ_{-i}\})=Y_i(\{Z_i=0,\bZ'_{-i}\})~~\forall \bZ \ne \bZ'$ and where $-i$ means "all vertices not $i$". $Y_{it}(D_{it})$ is the potential outcome of vertex $i$ at time $t$ under exposure condition $D_{it}$. The exposure condition indicates a mapping of the graph topology and $\bZ$ into the categories of exposure defined by the researcher (e.g., directly treated and not tied to any directly treated vertices, not directly treated and adjacent to at least one directly treated node).


\begin{table}[ht]
\centering
\begin{tabular}{rrrrrrr}
  \hline
 & A & B & C & D & E & F \\ 
  \hline
A & 0 & 1 & 1 & 0 & 0 & 0 \\ 
  B & 1 & 0 & 0 & 0 & 0 & 0 \\ 
  C & 1 & 0 & 0 & 0 & 0 & 1 \\ 
  D & 0 & 0 & 0 & 0 & 1 & 0 \\ 
  E & 0 & 0 & 0 & 1 & 0 & 0 \\ 
  F & 0 & 0 & 1 & 0 & 0 & 0 \\ 
   \hline
\end{tabular}
\caption{Adjacency matrix of a simple 6 node network with 4 edges.} 
\label{fig:toy-example-adj}
\end{table}To motivate this discussion, and introduce the primary statistical method, we introduce a toy example using a simpler model and test statistic than will be later employed. The primary goal of this toy example is demonstrate how the \emph{power}, the probability of rejecting a false null hypothesis, of the statistical approach can depend heavily on the treatment assignment mechanism, the network, and the interaction of these two features in generating the observed outcome. Table~\ref{fig:toy-example-adj} shows the adjacency matrix of this simple network. We label this network $S$. 
Additionally, for each node we have a background covariate $\bx = (2.4,0.6,2.2,0.9,0.4,0)'$. The covariate and the network combine with the treatment assignment $\bZ$ in two outcome models:
\begin{align}
  \bY(\bZ) &= \bx + \theta \bZ \label{eq:toy-no-spill} \\ 
  \bY(\bZ) &= \bx + \theta \bZ + (1 - \bZ) \theta I(S \bZ > 0) \label{eq:toy-no-spill2}
\end{align}
In the first model, as each node's outcome depends only on its own treatment status, this model involves no spillover. The second model, on the other hand, treated units share their benefit with any untreated neighbors. 
For simplicity, we fix the true $\theta$ at 1 in both models. To constrast different assignment mechanisms, we consider two treatment assignment mechanisms: assign two units to treatment, four to control (``unbalanced''); three units to treatment, three to control (``balanced''). For each possible treatment strategy, we can enumerate all possible treatment allocation (15 for the unbalanced case and 20 for the balanced case). For each allocation, we can generate the $\bY$ that would be observed if $\theta = 1$. For each of these observed $\bY$, we then apply the randomization inference procedure, using a squared t-statistic, to compute $p$-values for the (false) hypothesis that treatment had no effect (i.e., $\theta = 0$). This results in a $p$-value that would be observed for each possible randomization. For a given $\alpha$-level, the randomization method that has more $p$-values less than $\alpha$ has higher power.

\begin{figure}
  \centering
\includegraphics{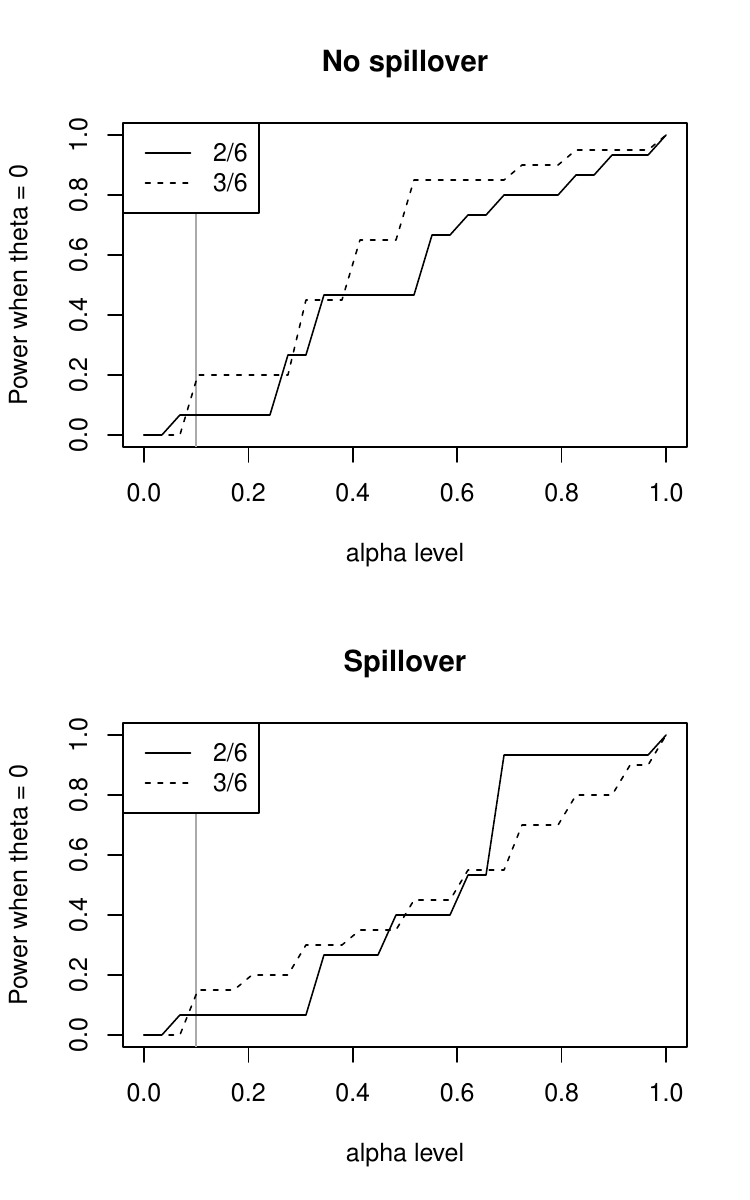}
\caption{\label{fig:toy-example-alpha-power} Proportion of randomizations that would reject the null hypothesis that $\theta = 0$ when $\theta = 1$ (``power'') of the two assignment methods (assigning 2 out of 6 or 3 out of 6 subjects to treatment) as a function of the $\alpha$-level employed. The top panel represent the case of the model of direct effects only, while the lower panel adds an indirect effect for any node that has at least one treated neighbor. The vertical gray line is at $\alpha = 0.10$}.
\end{figure}

Figure~\ref{fig:toy-example-alpha-power} compares the two assignment methods when spillover is and is not present with respect to power, as a function of the selected $\alpha$-level. As the figure shows, in the absence of spillover, the balanced design almost always possess better power characteristics than the unbalanced design. As power is a function of $p$-values generated for each randomization, it can be useful to consider the mean $p$-value of the sharp null hypotheses for these different methods, where lower mean $p$-values correspond to more power. Overall, for the balanced design the average $p$-value without spillover was $0.4$ while for the unbalanced design had mean $p$-values of $0.493$. Conversely, when spillover is present the unbalanced design generally has superior power to the balanced design, with an overall mean of  $0.547$ compared to $0.56$ for the balanced design. While the difference between the $p$-values for the spillover model is small, it is important to note that conventional wisdom holds that balanced designs always possess superior power advantages. When spillover is present, however, this is not universally true. In some situations, when spillover is present, unbalanced assignment can increase statistical efficiency.

To investigate this phenomenon in a more realistic setting, in the subsequent simulations we employ a much richer model of treatment effects as well as test statistics that have proven themselves useful in network settings. The model we use for treatment propagation in our simulations is a variant of the Ising Model. The initial treatment assignment of each vertex is drawn independently from $Z_i \sim \text{Bernoulli}(\alpha)$. The infection probability---the probability that treatment ``infects,'' or propagates to, a control vertex---at each iteration of propagation is
  $$\frac{1}{1+\exp(\frac{2}{F}(k_i-2m_i))},$$
where $k$ is number of (directly adjacent) neighbors, $m$ is the number of previously exposed neighbors ($0 \le m_i \ge M$), $F$ is a ``temperature'' parameter that governs the extent to which the propensity to be infected depends on the infection rate among $i$'s neighbors.  We initialize the model by assigning treatment ($t=0$), and then we run the propagation model for just one time period ($t=1$). The Ising model controls actual infection after an experimenter assigns $Z_i$ at $t=0$. By only producing one iteration of the model, we thus only allow interference at one step or degree in the network.

We specify the potential outcomes of the vertices in our simulation according to the following scheme depending upon the infection status at a given time point. We denote this by $Y(Z_{i,t=0},Z_{i,t=1})$, in which $Z_{i0} \equiv Z_{i,t=0}$ indicates $i$'s initial treatment status and $Z_{i1}$ indicates whether treatment propagates to $i$ at time 1. In the simulation study we consider both multiplicative and additive effects, and propagation models in which treatment propagates stochastically according to the Ising model, and propagates with certainty from directly treated units to their direct neighbors. We generate a  baseline (pre-treatment) response $ Y(0,0) \sim U(0,1)$ to represent the state of the graph in the absence of any experiment.  Our multiplicative treatment effect model changes the baseline in the same, multiplicative, way regardless of the time or manner of ``infection'' (directly assigned by researcher or propagated from a neighbor), and $Y(1,0) = Y(0,1) =  \lambda Y(0,0)$.  In the additive treatment effect model $Y(1,0) = Y(0,1) =  \lambda + Y(0,0)$. We consider values of $\lambda \in \{0.26, 0.63\}$, which correspond to approximately one and two standard deviation shifts in the mean of a standard uniform random variable, and simulate 1,000 treatment propagation and outcome sets at each combination of $F \in \{0,10,\hdots,100\}$ and $\alpha \in \{0.05, 0.10,\hdots,0.50\}$.

  \subsection{Application of Inference Methods}

  \citeapos{aronow2017interference} method requires that we  define exposure conditions (which
include assignment to treatment and also the probability of exposure to a
treatment via the network). We define the exposure conditions with respect to
what a researcher would be able to observe from the experimental design,
assuming the network were observed, and the response. Importantly, in defining
the exposure conditions of interest, we assume that the researcher does not
exactly know the set of vertices to which the treatment has propagated. We see
this as more realistic than a situation in which the researcher knows exactly
where the treatment has propagated.

\begin{figure}[!ht]
\begin{center}
\includegraphics[width=.7\textwidth]{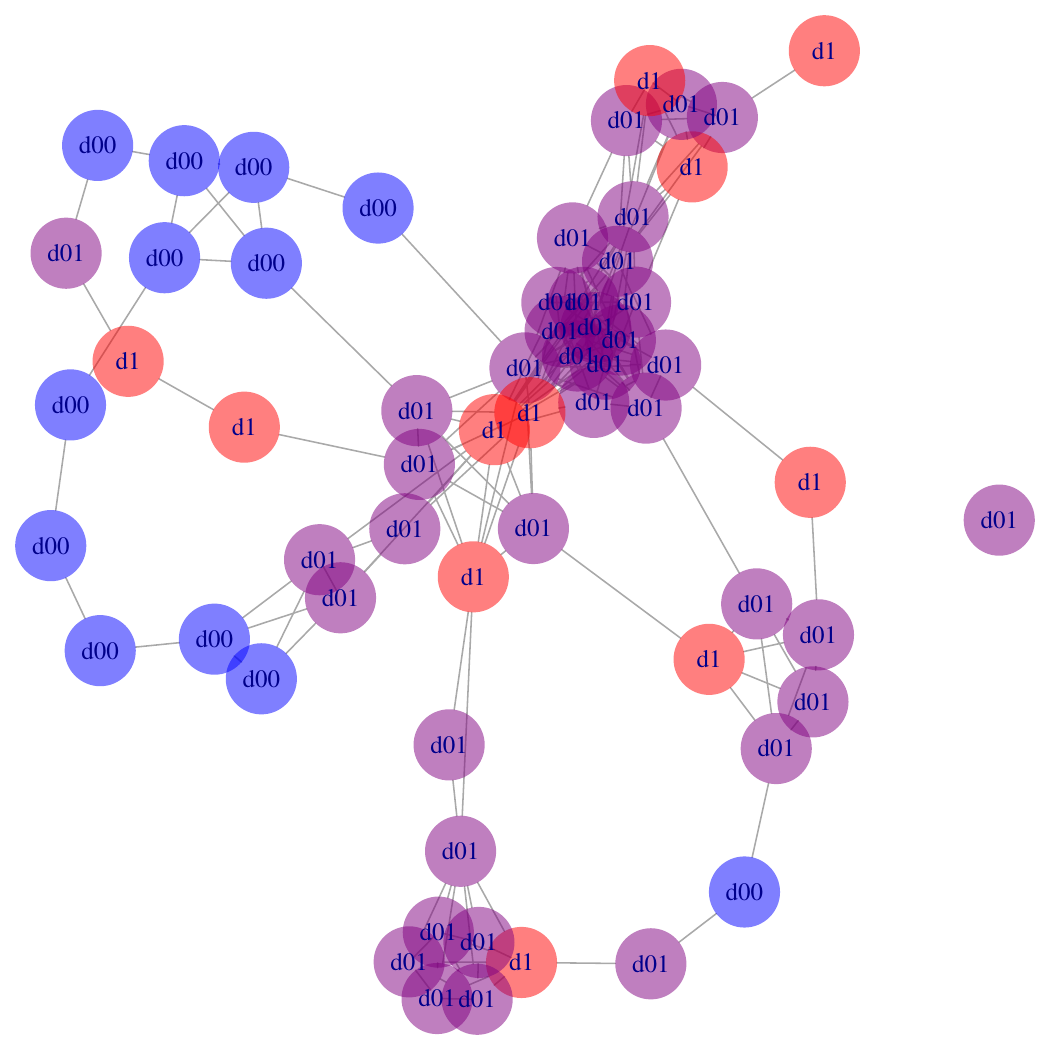}
\end{center}
\caption{Illustration of exposure conditions for a subgraph of the Ghana road
  network.  One draw from the Ising propagations with  $\text{pr}(Z_i=1) \sim \text{Bernoulli}(.15)$
  for $t \in \{0,1\}$ and Temperature$=10$.}
\label{fig:exposure}
\end{figure}

We define the following three distinct exposure conditions, differentiating between those vertices treated initially ($d_1$), those vertices that are untreated initially and are adjacent to at least one treated vertex ($d_{(0,1)}$), and those vertices that are untreated initially and are not adjacent to any treated vertices ($d_{(0,0)}$).
 \begin{itemize}
  	    \item $d_1 \equiv D_i(Z_i=1,0 \le m_{it} \ge M)$
	    \item $d_{(0,0)} \equiv D_i(Z_i=0,m_{it}=0)$
	    \item $d_{(0,1)} \equiv D_i(Z_i=0,m_{it} \ge 1)$
	    \end{itemize}
Figure \ref{fig:exposure} gives a visual example of a subgraph drawn from the Ghana road network. In the study of the \citet{aronow2017interference} methods, we focus on identifying the difference between the $d_{(0,1)}$ and $d_{(0,0)}$ conditions.

The \citet{aronow2017interference} estimand is the difference between vertices in two exposure conditions. First, let
  $$ \hat{\mu}(d_k) = \frac{1}{n} \sum_{i=1}^n \text{\bf I}(D_i = d_k) \frac{Y_i(d_k)}{\pi_i(d_k)},$$
 be the estimator of the mean potential outcome among vertices in exposure condition $d_k$, where $\pi_i(d_k)$ is the probability that vertex $i$ ends up in condition $d_k$. Then the estimator of the difference between potential outcomes in the two exposure conditions is

  $$ \hat{\tau}(d_{(0,1)}, d_{(0,0)}) = \hat{\mu}(d_{(0,1)})- \hat{\mu}(d_{(0,0)}).$$

Following Horvitz and Thompson (1952), \citet{aronow2017interference} show that \\
  $\text{Var}( \hat{\tau}(d_{(0,1)}, d_{(0,0)}) ) \le 1/N^2 \left( \text{Var}(\hat{\mu}(d_{(0,1)})) + \text{Var}( \hat{\mu}(d_{(0,0)})) \right) $. In the interest of brevity, we do not re-produce the full variance estimators, but refer readers to \citet{aronow2017interference}. A power analysis gauges the probability that a hypothesis (usually the null hypothesis) is rejected when it is false. In the results that follow, we assess the power to reject $H_0: \tau=0$ at the 0.05 significance level.

 We test hypotheses about effects following  \citet{Bowers:2013} using the Anderson-Darling $k$-sample test statistic \citep{Scholz1987} to compare the outcome distributions of vertices in the three different exposure conditions.\footnote{Randomization distributions are simulated using a development version of the \texttt{RItools} package for \R \citep{Bowers2014}.} Following the norms of assessing power against truth, we test hypotheses generated by the correct Ising model and record the amount of false rejections as the parameter values of $\lambda$ and $F$ move away from their true values. The values of the parameters $\lambda$ and $F$ are manipulated to test the null of no effects (i.e., $\lambda =1, F = \infty$) in order to assess power. In the case of the methods proposed by \citet{Bowers:2013}, we assess power with respect to two different null hypotheses. The first is the null of no effects across all three exposure conditions (i.e., the null that the experiment did not effect any vertices). The second null hypothesis is the null of no difference between $d_{(0,1)}$ and $d_{(0,0)}$ (i.e., the null that there is no interference in that there is no difference between isolated controls and controls that are adjacent to treated vertices).

\subsection{Results}

\begin{figure}[H]
\begin{center} \vspace{-.7cm}
    \begin{tabular}{cc}
      {$\lambda$ = 0.26} & {$\lambda$ = 0.63} \\
            \multicolumn{2}{c}{(a) Multiplicative Effects  ($Y(1,0) = Y(0,1) = \lambda Y(0,0)$)}\\
      \includegraphics[scale=.575]{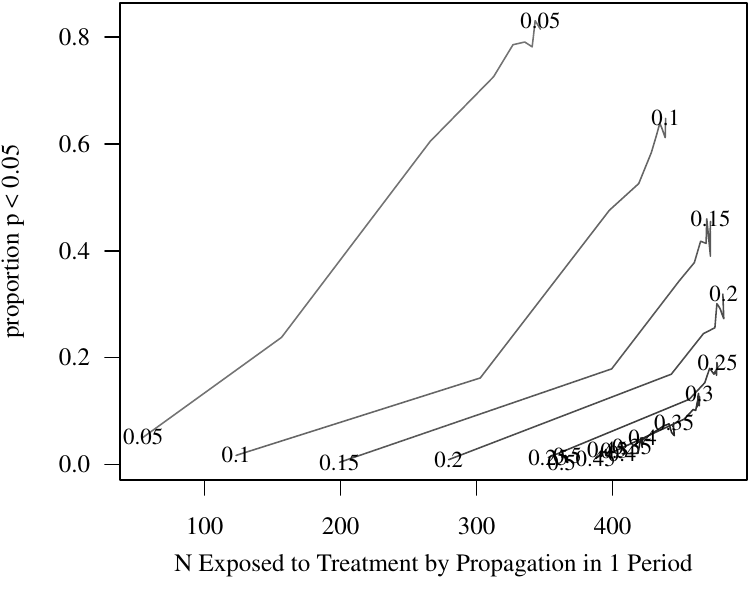} & \includegraphics[scale=.575]{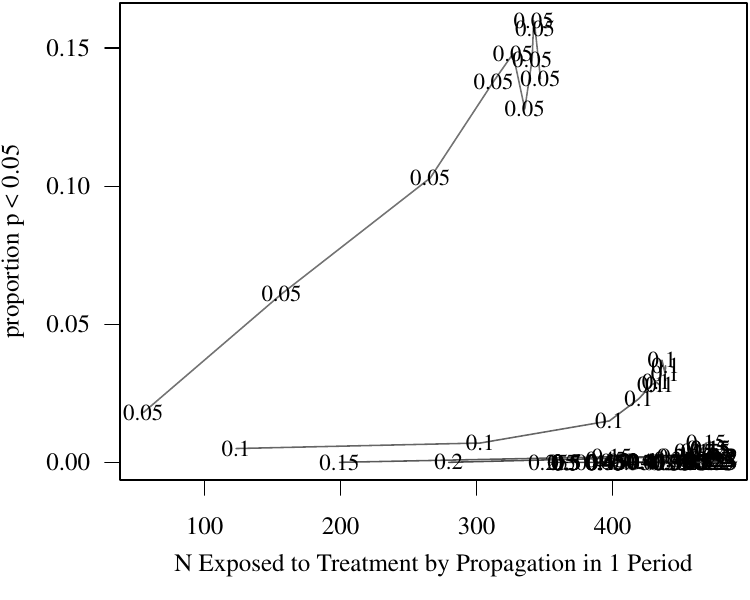}\\
      \multicolumn{2}{c}{(b) Additive Effects  ($Y(1,0) = Y(0,1) = \lambda + Y(0,0)$)}\\
      \includegraphics[scale=.575]{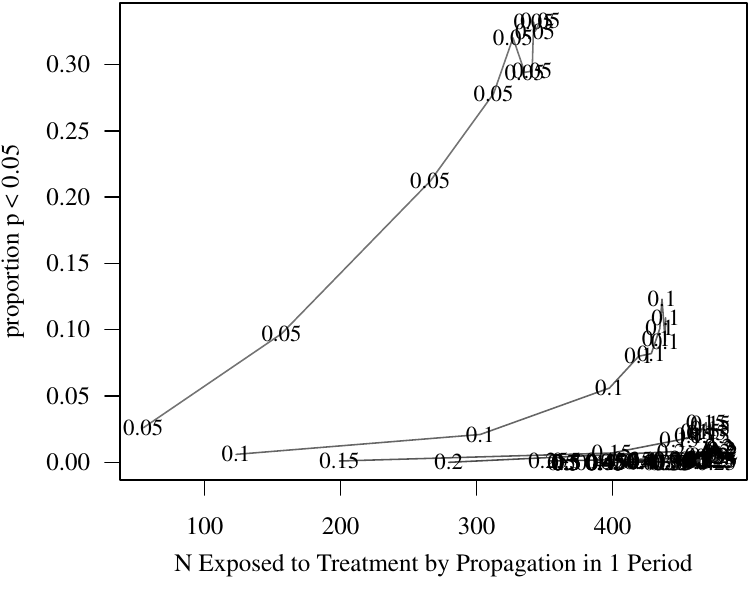} & \includegraphics[scale=.575]{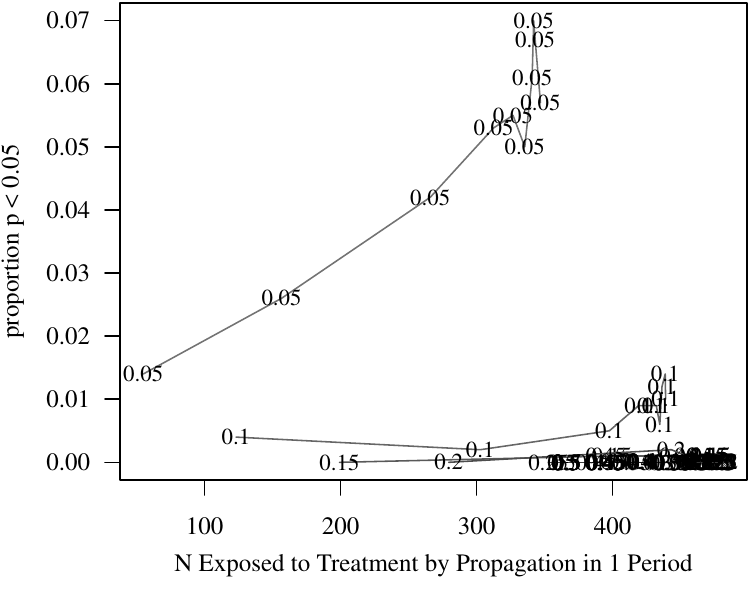}\\
        \multicolumn{2}{c}{(c) Perfect Propagation ( $Y(0,1) = Y(d_{0,0})$)}\\
      \includegraphics[scale=.575]{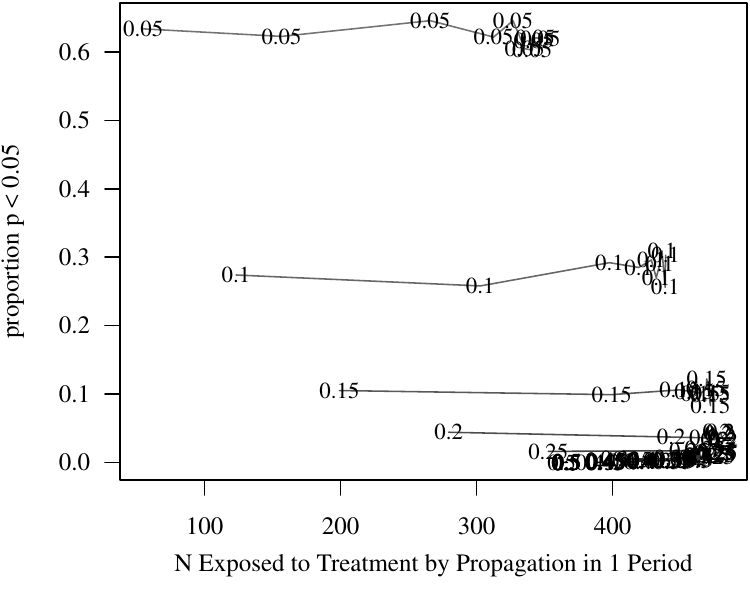} & \includegraphics[scale=.575]{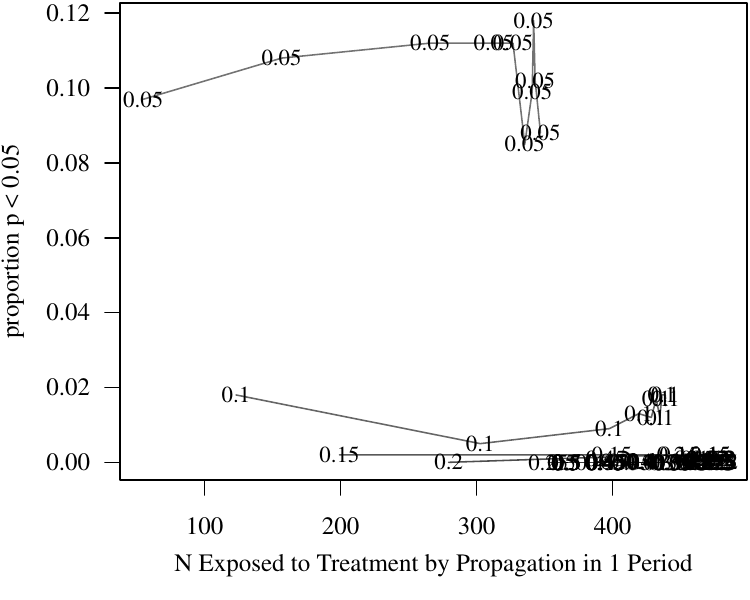}\\
    \end{tabular}
    \end{center} \vspace{-1cm}
    \caption{Power Results: \citet{aronow2017interference} test. Lines are labeled by the proportion assigned to treatment. High power depends on low proportions assigned to treatment. The network contains a total of 868 nodes. The $y$-axes show that nominal power of .8 is not always attained in all designs.}
    \label{fig:aranowsamiipower}
\end{figure}

The results from the \citet{aronow2017interference} tests are reported in Figure \ref{fig:aranowsamiipower}. Three distinct sets of results are presented: (a) those with a multiplicative effect and stochastic propagation (i.e., $Y(1,0) = Y(0,1) = \lambda Y(0,0)$), (b) those with an additive effect (i.e., $Y(1,0) = Y(0,1) = \lambda + Y(0,0)$), and (c) those with an additive effect and certain propagation (i.e., $Y(0,1)= Y(d_{(0,1)})$). We note three characteristics of our results. First, the tests exhibit fairly low power overall, hitting a maximum of approximately 0.8, but sitting below and often far below 0.5.\footnote{It is notable that \citet{aronow2017interference} present an alternative test, based on a Hajek estimator that they show is more efficient than the HT estimator (with, however, some bias). We present the results of this simulation using the Hajek estimator in the Appendix. We see that the Hajek estimator does exhibit higher power, and that we see the same patterns in terms of the proportion assigned to treatment. However, we report the HT estimator in the main text, as some researchers may prefer to use the unbiased estimator, and, more practically speaking, the Hajek exhibits such high power that it is more difficult to see how power varies with the conditions in the simulation study using the Hajek estimator, as compared to the HT estimator. } Second, the most powerful design is that in which $\alpha=0.05$, the smallest proportion assigned to initial treatment.  Third, as indicated by the relationship of power with the $x$-axis of the plots in panels (a) and (b), the larger the sample of vertices in the $d_{(0,1)}$ condition that are actually exposed to treatment in period one, as governed by the temperature parameter, the more powerful the tests.

\begin{figure}[H]
\begin{center}
(a) Treating High Centrality Nodes \\ \vspace{-.3cm}
 \includegraphics[scale=.9,clip=true,trim=3cm 2cm 2cm 2cm]{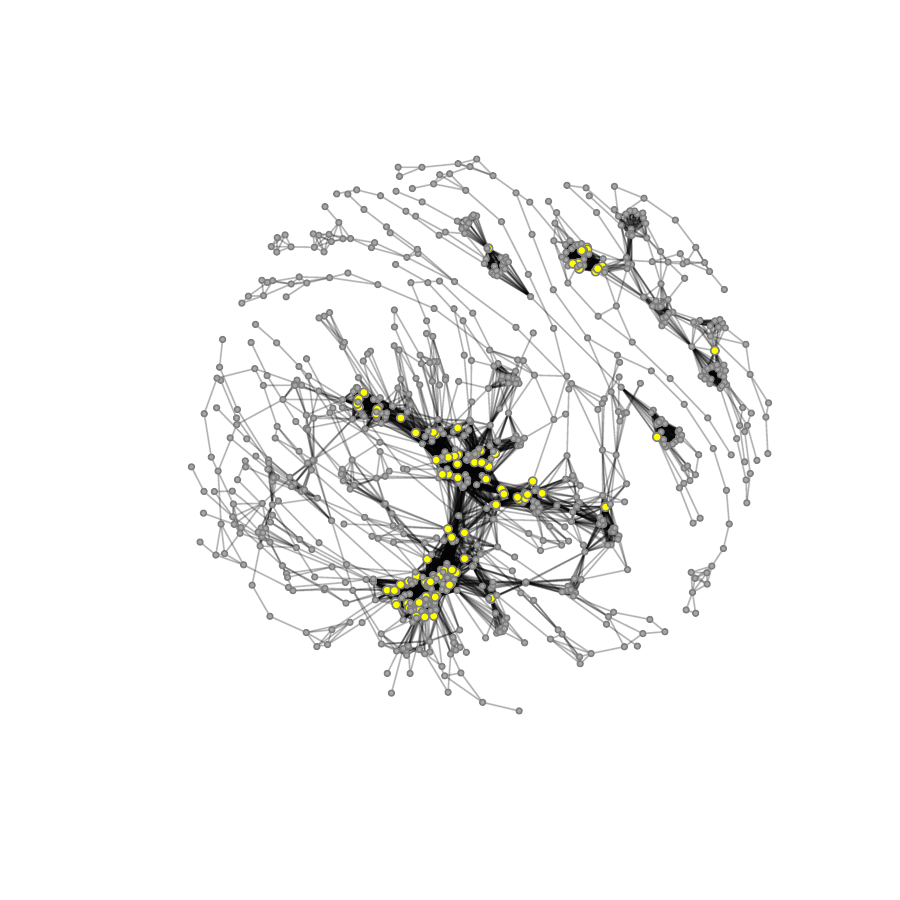} \\  \vspace{-1cm}
 (b) Treating Low Centrality Nodes \\ \vspace{-.3cm}
  \includegraphics[scale=.9,clip=true,trim=3cm 2cm 2cm 2cm]{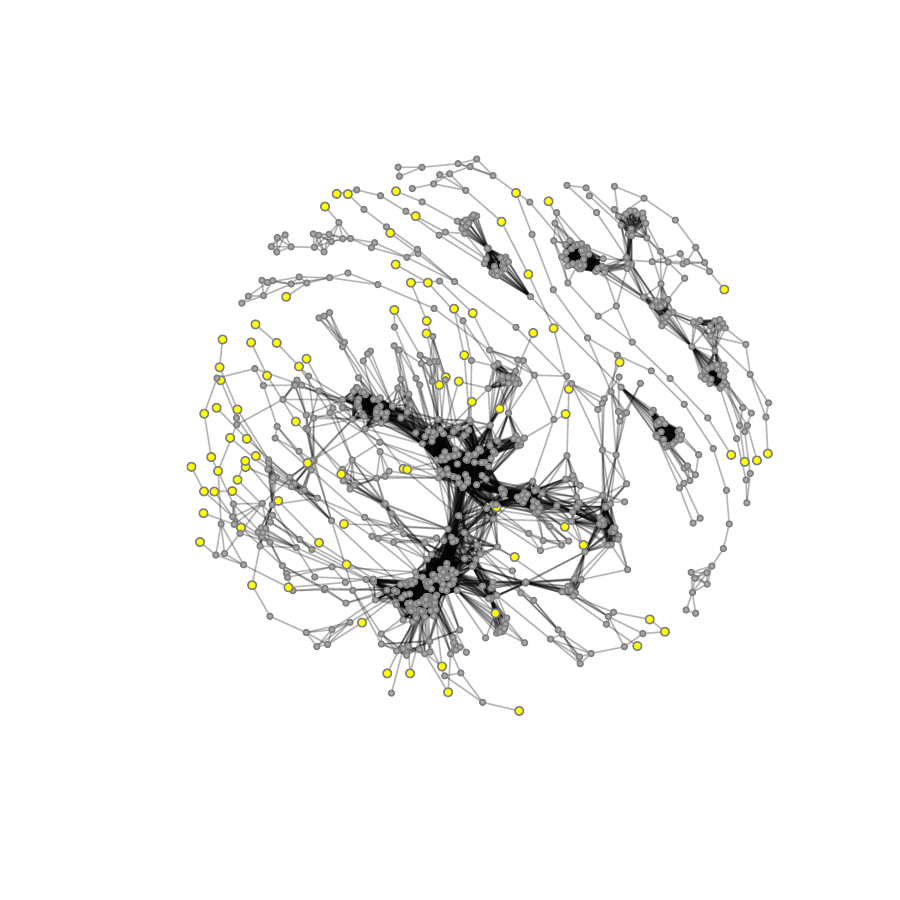} \\
 \vspace{-1.5cm}
  \end{center}
 \caption{Examples of treatment assignment to either high or low degree nodes in the Ghana 2008 Voter Registration Fraud Experiment network. Vertices are registration stations and yellow nodes are assigned to initial treatment.}
 \label{fig:networks}
\end{figure}

\begin{figure}[H]
\begin{center}
\includegraphics[scale=.95]{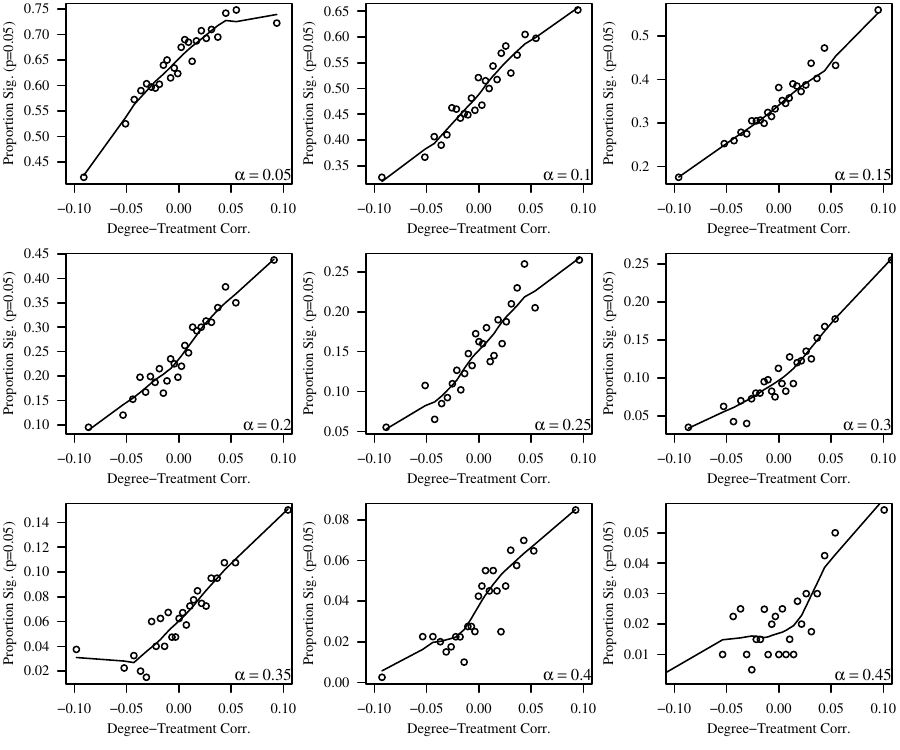}
\end{center}
\caption{When node degree is correlated with probability of treatment assignment, the power to detect indirect effects increases. Results are drawn from the simulation condition with the multiplicative model of effects, and stochastic propagation of treatment.}
 \label{fig:degcor}
\end{figure}

If a researcher wants to estimate the network exposure weighted average causal effect developed by  \citet{aronow2017interference} these kinds of results raise the question about whether the best approach to randomization of treatment assignment is simple uniform assignment without any blocking or use of information about the fixed network. To investigate this we study how power depends on the correlation between $m$ and $Z_{i0}$ (i.e., the correlation between vertex degree and the initial treatment assignment) in the simulation condition with the multiplicative effects and stochastic propagation. The networks depicted in Figure \ref{fig:networks} represent examples of treatment assignments biased in favor of high degree vertices (a) and low degree vertices (b). The degree-treatment correlation results are given in Figure \ref{fig:degcor}. We can see in Figure \ref{fig:degcor} that at each $\alpha$ value there is a strong positive relationship between the degree-treatment correlation and statistical power. This indicates that for the Ghana network structure and the propagation/effects models we have specified, designs that bias treatment assignment towards higher degree nodes would exhibit greater statistical power. This finding generalizes what we know about the use of prognostic background covariates to increase power in randomized experiments to the situation where network degree can be thought of as a moderator of treatment effects.

\begin{figure}[!ht]
    \begin{center}
     \includegraphics[width=.7\textwidth]{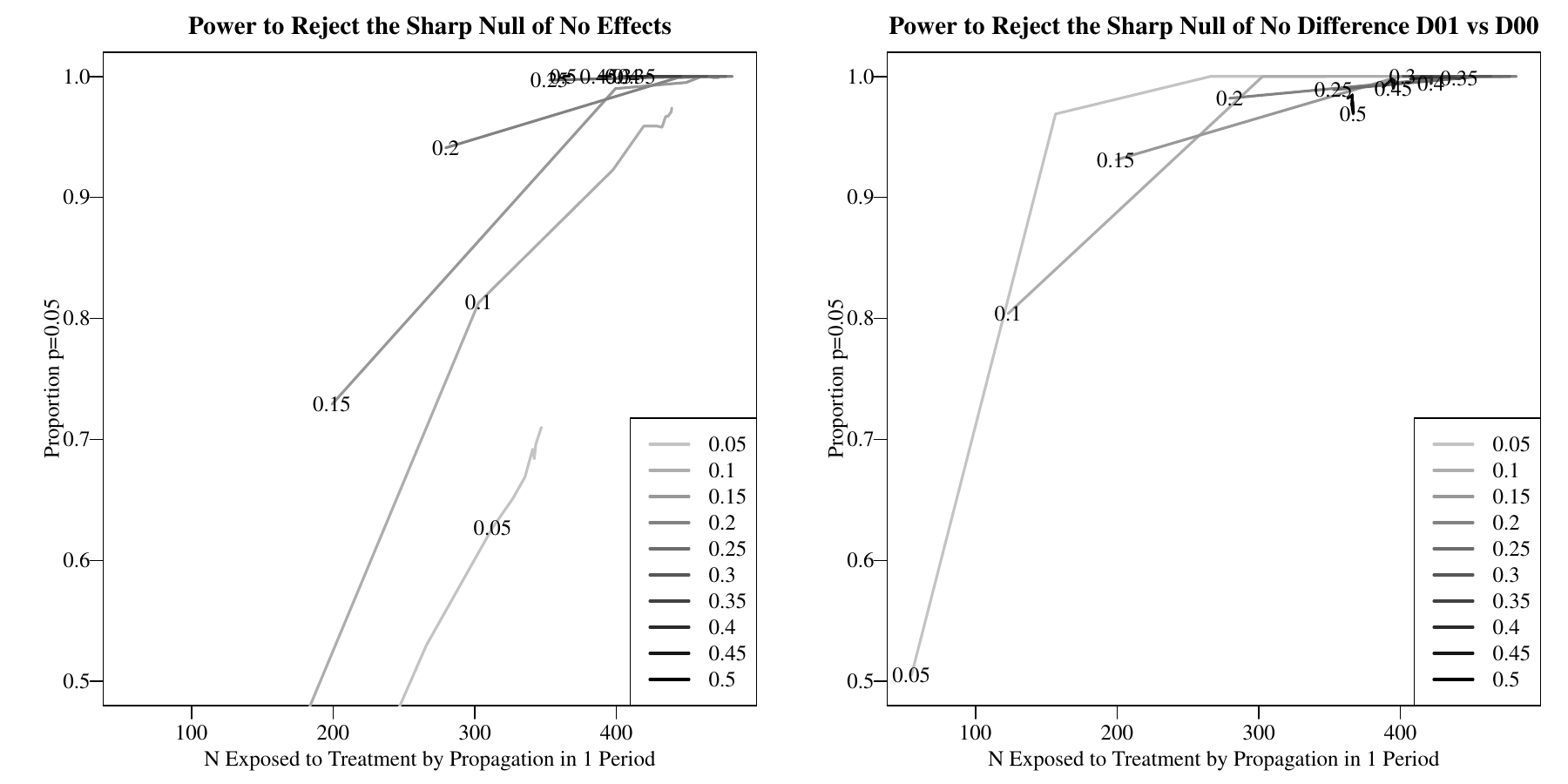}
     \end{center}
     \caption{Power with multiplicative effects model and true Ising propagation model following  \citet{Bowers:2013}.}
     \label{fig:bowersmult}
    \end{figure}

\begin{figure}[!ht]
    \begin{center}
     \includegraphics[width=.7\textwidth]{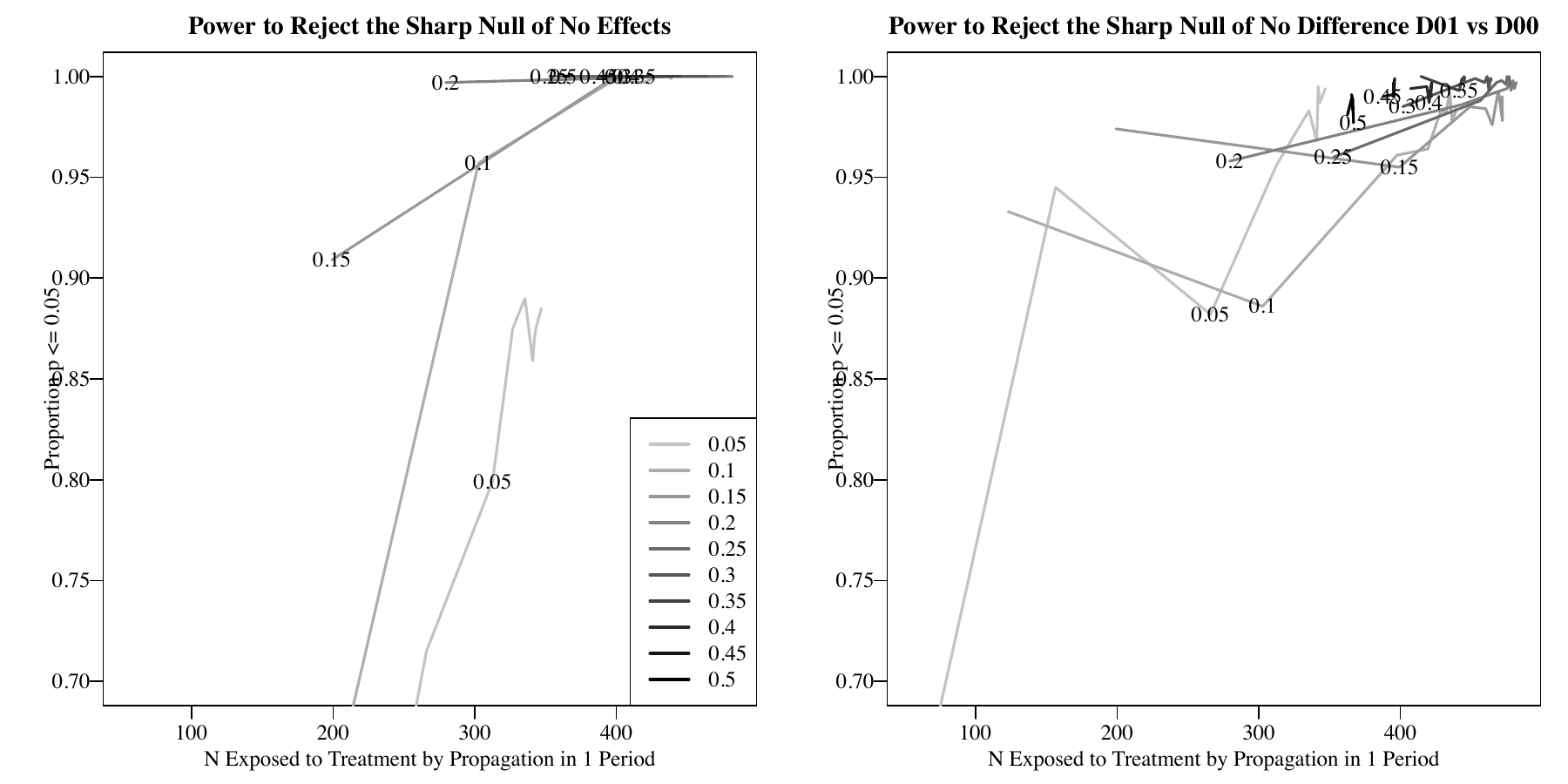}
     \end{center}
     \caption{Power with additive effects model and true Ising propagation model following  \citet{Bowers:2013}.}
     \label{fig:bowersadd}
    \end{figure}

\begin{figure}[!ht]
    \begin{center}
     \includegraphics[width=.7\textwidth]{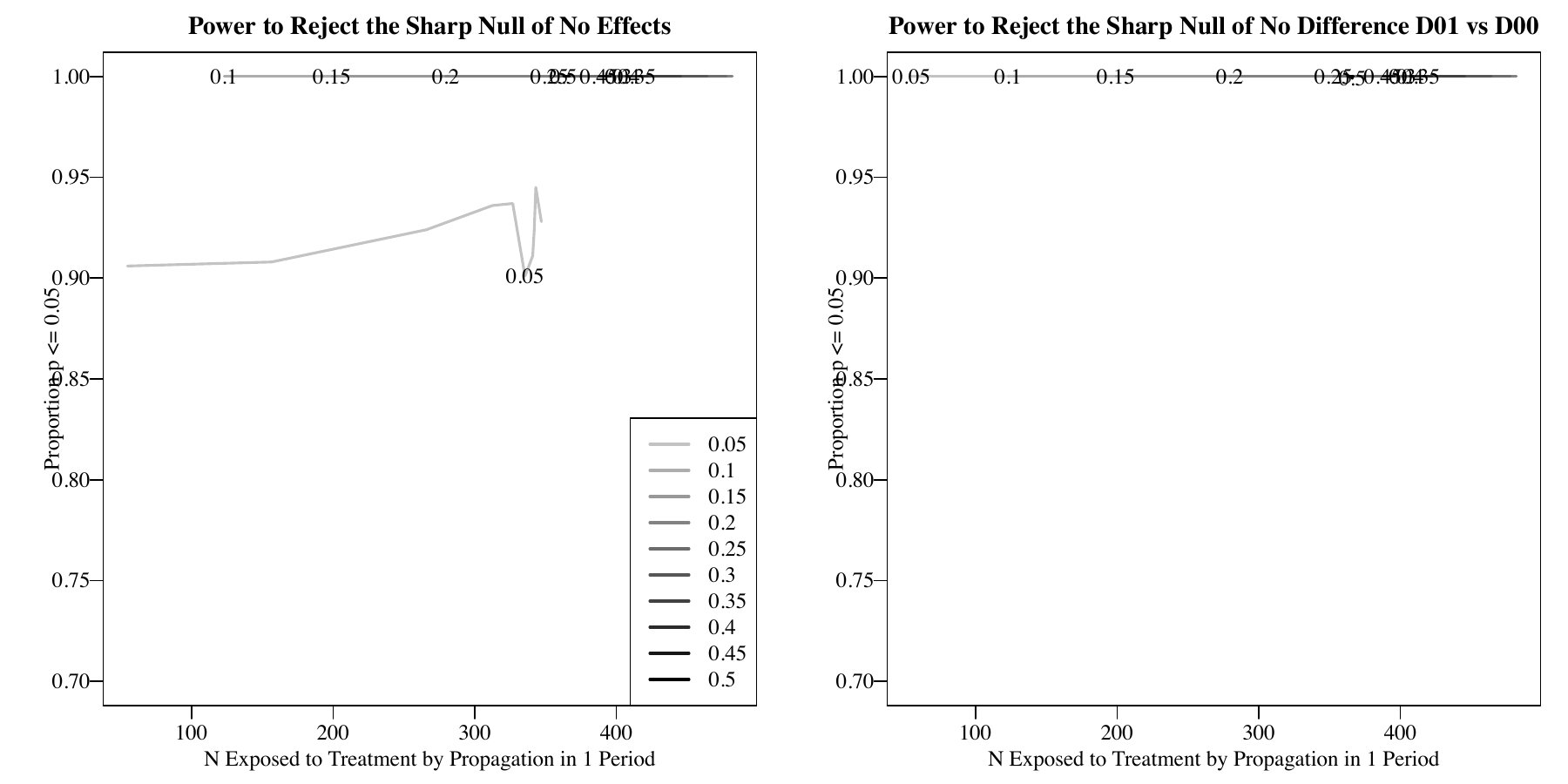}
     \end{center}
     \caption{Power with additive effects and perfect propagation and true Ising propagation model following  \citet{Bowers:2013}.}
     \label{fig:bowerscert}
    \end{figure}

The results from the model-testing methods of \citet{Bowers:2013} in Figures \ref{fig:bowersmult}--\ref{fig:bowerscert} differ in their implications from the results based on estimation. First, the power in these tests is much higher overall. This makes some sense: we are assessing the ability of the test to reject false parameters given a correct propagation model, a good test should eventually reject values of parameters that are distant from the truth. The methods focused on differences of averages do not include much information about the propagation model except as a weight arising indirectly from what we can observe, so those procedures have less information to use in statistical inference in this simulation study.  Second, when we focus on the test of the sharp null of no effects when all three exposure conditions are included, we see that the low values of $\alpha$ exhibit low power. However, when the test  compares only $d_{(0,1)}$ against $d_{(0,0)}$, which can be considered a test for propagation effects (i.e., excluding directly treated vertices), the lower $\alpha$ designs perform better. This tradeoff between the power to detect any effect versus the power to detect propagated effects replicate the findings from  \citet{Bowers:2013}: in order to learn about propagation, one should assign relatively few nodes to treatment, in order to test an overall null of no effects, then more power arises from more directly assigned-to-treatment nodes.  There is also a positive association between the number of vertices in the $d_{(0,1)}$ condition that are exposed to treatment by period one and the power of the tests. This last result arises from the fact that, when few in the $d_{(0,1)}$ condition are exposed to treatment, the $d_{(0,1)}$ outcomes are, in large part, equivalent to the $d_{(0,0)}$ outcomes.

    \clearpage

    \subsection{Summary}

    We have shown that learning about a simple model of propagation of
    treatment through a network (via comparisons of nodes possibly indirectly
    exposed to the treatment with nodes not exposed to treatment) can be enhanced
    when relatively few of the nodes in a network are assigned treatment. We also find, at least with the Aronow and Samii method, that power may be improved by using network structure in formulating the randomization distribution (e.g., assigning treatment to vertices with probability proportional to their degrees in the network).  Both the Bowers et al. method and the Aronow and Samii method explicitly account for the randomization distribution, and we are, as such, not concerned about introducing bias into the methodology by biasing the randomization distribution. This would be a particularly important method if the researcher hypothesized that the model of effects varied with respect to network structure (e.g., a hypothesis that higher degree vertices are more susceptible to the treatment). Since the randomization distribution is incorporated into the methods we present, both approaches could be used to test hypotheses regarding effects that vary with respect to network structure. More abstractly,
    our analysis acts an example for those desiring to evaluate their own
    models of propagation on networks and experimental designs before going
    into the field. Indeed, our findings that the power-maximizing proportion assigned to treatment falls well below 0.50, and power can be increased by disproportionately assigning treatment to higher degree nodes is specific to the network dynamics that we artificially designed for this simulation study (i.e., the Ghana road network structure, the Ising model for treatment propagation, and our model(s) of effects). However, simulation studies such as the ones we have run would assist the researcher in optimally designing the experiment given a network structure, and model(s) that characterized propagation and treatment effects. We do not consider our findings regarding the proportion assigned to treatment and disproportionate assignment to highly connected nodes to apply universally to experiments on networks.

    We have not shown how the \citet{Bowers:2013} approach
    performs when an incorrect model is assessed. We did not do this because we
    are focusing on design---and power analysis requires that we create a
    truth against which to compare alternatives. \citet{Bowers:2013, bowers2016research} show that,
    in the analysis stage, hypothesis tests may have very low or no power if
    the model being assessed has no bearing on the underlying mechanism, but we
    are not certain what kind of design advice would follow from such findings---merely increasing the size of a fixed network may be impossible.

  We have also not considered, for either the Aronow and Samii or Bowers et al.\ methods, how design affects statistical power when there is uncertainty regarding the network structure. To incorporate uncertainty regarding the network, a stochastic model for the network would need to be integrated into the analytical procedure(s) \citep[e.g., ][]{desmarais2012}. Relatedly, network-based sampling methods (e.g., snowball sampling) are commonly used to study hard-to-reach populations (e.g., illicit drug users \citep{wang2007respondent}). If researchers are introducing interventions in network-based samples, the effects of the interventions may include interference. There is an active literature in statistics and computer science that considers the ways in which network quantities (e.g., prevalence of a behavior, degree distribution) can be accurately and efficiently estimated via network sampling designs \citep{handcock2010modeling,kurant2012coarse,gile2010respondent}.  We have not thoroughly considered how the tests studied in the current paper, and associated approaches to treatment assignment, would perform in the context of network sampling, but future work should consider the integration of tests for and estimates of interference effects with network sampling designs. Such consideration would involve the assessment of optimal treatment assignment distributions (e.g., disproportionately treating higher degree nodes) based on noisy information regarding network structure, gathered through a network sampling design.

\section{Conclusion}

We describe the challenges in experimental design that arise when the researcher is interested in studying the process of propagation on a network with the objective of drawing causal inference. The experimental designs that work most effectively in experiments in which there is no interference are unlikely to be directly transferable to experimental research on propagation. We review two recently developed frameworks for statistical inference regarding interference in networks. One commonality we draw from these two frameworks is that theoretical analytic models of propagation play a key role in their application, which means that substantive theory about the nature of treatment effects and network relations features more prominently in the statistical analysis of experimental data generated for the study of propagation than in the classical, non-interference, experimental framework.

We present a simulation study to (1) illustrate how simulation can be a useful guide in identifying design parameters for experimental studies of interference, and (2) study the properties of the two frameworks for statistical inference presented in the front end of the paper. Three findings from the simulation study are notable. First, statistical power depends upon design parameters and, for example, the optimal proportion assigned to initial treatment may be much lower than the conventionally applied 0.5.\footnote{This finding is supported by \citet{baird2017optimal}'s study of partial intereference,  networks where the nodes are isolated into subsets. They show similar results in regards the tradeoffs between power to detect direct and peer-effects while focusing on on a version of the Aronow and Samii estimator tailored for their specific design.} Second,  the relationship between design parameters and power depends upon the framework for statistical inference. Third, our results regarding the positive relationship between the degree-treatment correlation and power indicates that randomization designs that take into account graph topology are likely to exhibit substantial power gains over uniform randomization designs. It is important to note, however, that these findings are specific to our simulation setup, and may not apply directly to other network experiments defined by different networks, models of propagation, and/or models of treatment effects. Nevertheless, the results from the simulation study underscore the importance of considering design parameters for experimental studies of propagation, as the standards of the classical experimental framework are unlikely to apply. We encourage the use of such simulation studies to guide the design of experiments on networks.

Our results suggest three fruitful directions for future research. First, as noted above, the literature on network dynamics offers several possibilities for specifying models of propagation on networks. Researchers may not have strong a priori theory regarding the functional form of the propagation model.  This leads to the first future direction---considering whether the propagation model can be learned algorithmically, or analyzed though a nonparametric framework.  Second, although we focused on power to detect indirect and direct effects, we were studying multiple hypotheses (two in this case). Our work here has made us wonder whether designs to maximize power against \emph{combined tests} of those (and other) hypotheses might look different from designs which aim only at maximizing power to detect propagation or overall effects.  The third question regards the network through which interference occurs. Future work should consider precisely how uncertainty regarding the network structure can be incorporated into the methods we present.



\bibliographystyle{apa}
\bibliography{references}

\section*{Appendix}

  \begin{figure}[H]
\begin{center} \vspace{-.7cm}
    \begin{tabular}{cc}
      {$\lambda$ = 0.26} & {$\lambda$ = 0.63} \\
            \multicolumn{2}{c}{(a) Multiplicative Effects  ($Y(1,0) = Y(0,1) = \lambda Y(0,0)$)}\\
      \includegraphics[scale=.575]{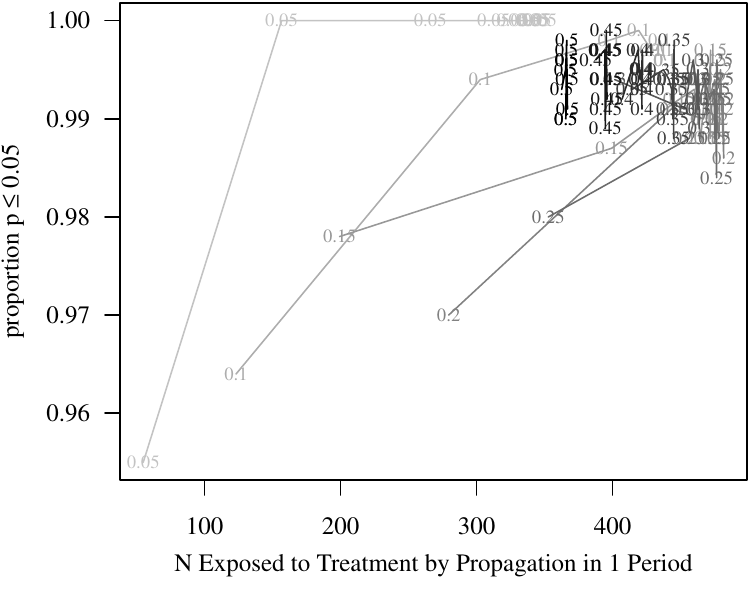} & \includegraphics[scale=.575]{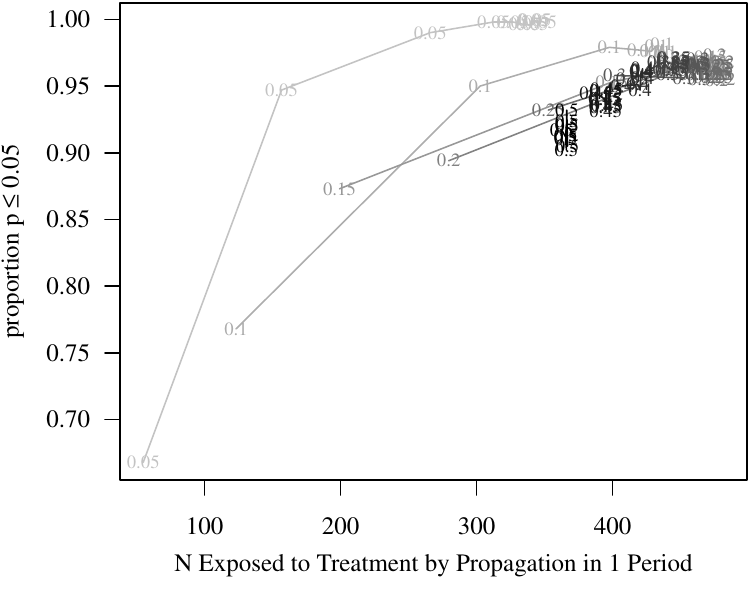}\\
      \multicolumn{2}{c}{(b) Additive Effects  ($Y(1,0) = Y(0,1) = \lambda + Y(0,0)$)}\\
      \includegraphics[scale=.575]{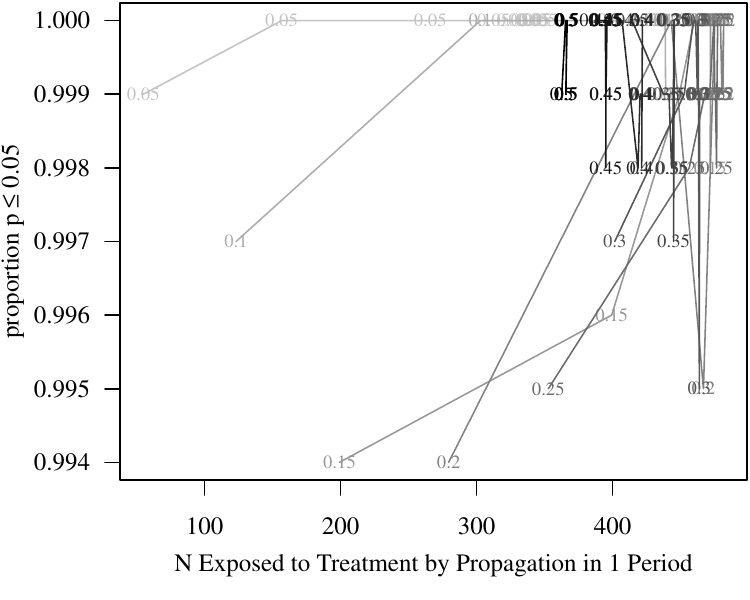} & \includegraphics[scale=.575]{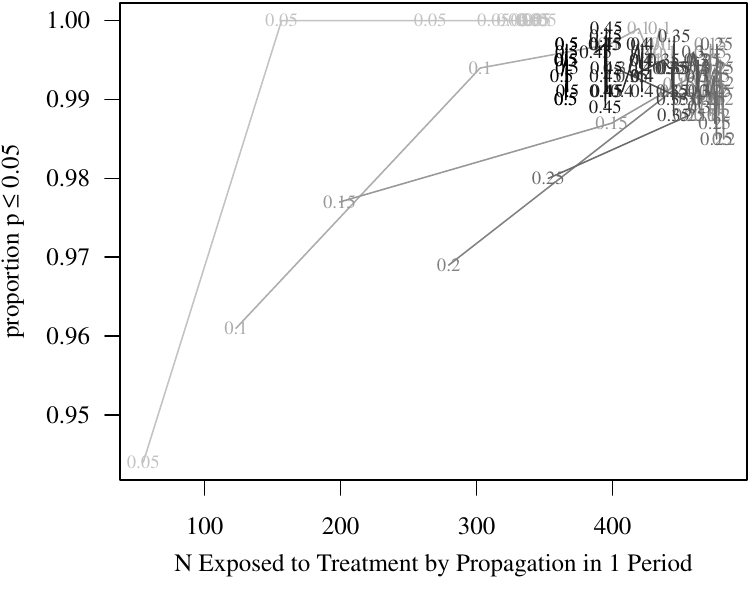}\\
        \multicolumn{2}{c}{(c) Perfect Propagation ( $Y(0,1) = Y(d_{0,0})$)}\\
      \includegraphics[scale=.575]{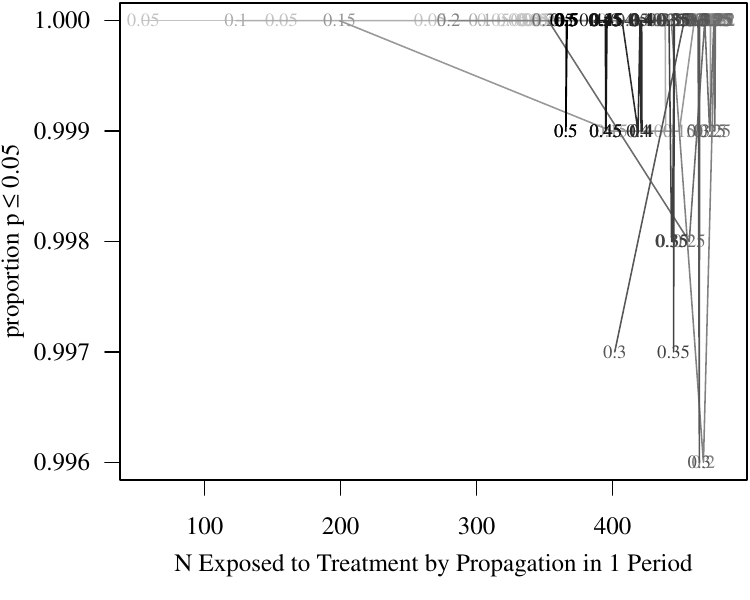} & \includegraphics[scale=.575]{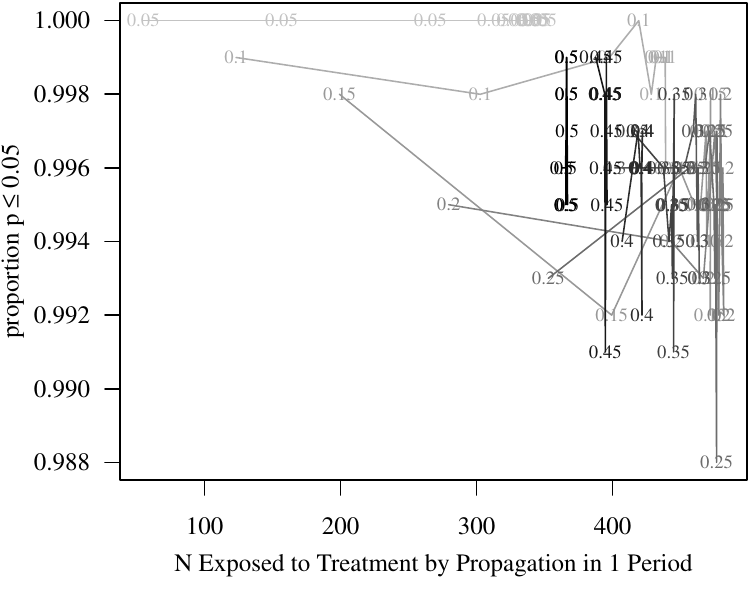}\\
    \end{tabular}
    \end{center} \vspace{-1cm}
    \caption{Power Results: \citet{aronow2017interference} (Hajek) test. Lines are labeled by the proportion assigned to treatment. High power depends on low proportions assigned to treatment. The network contains a total of 868 nodes. The $y$-axes show that the test based on the Hajek estimator exhibits much higher power than the test based on the HT estimator.}
    \label{fig:aranowsamiipowerHajek}
\end{figure}

\end{document}